# Dual-Wavelength Bi-Doped Fiber Laser

# Based on Cascaded Cavities

**Galina Nemova,[a] Xian Jin,[a] Lawrence R. Chen,[a] Sergei V. Firstov,[b] and Omur Sezerman[c]**

[a]Department of Electrical and Computer Engineering, McGill University,
McConnell Engineering Building, 3480 University Street, Montreal, QC, H3A 0E9 Canada
[b]Fiber Optics Research Center of the Russian Academy of Sciences, 38 Vavilov Str., Moscow 119333, Russia
[c]OZ Optics, 219 Westbrook Rd, Ottawa, ON, K0A 1L0 Canada

**Abstract:** We present a comprehensive theoretical and experimental investigation of a dual-wavelength, bismuth-doped fiber (BDF) laser operating near 1700 nm based on cascaded cavities. The BDF provides optical gain from 1650 nm to 1800 nm when pumped at 1550 nm; the laser cavity is defined by a 90% mirror on one end and two fiber Bragg gratings (FBGs) separated by a length of BDF on the other. The laser can operate at either wavelength alone or both wavelengths simultaneously by simple adjustment of the pump power. Experimental results correlate well simulations based on a theoretical model of the laser.

## 1. Introduction

The first observation of infrared (IR) photoluminescence in bismuth (Bi) doped glasses was announced in the late 1990's [1]. In 2005, bismuth-doped fibers (BDFs) were fabricated by the modified chemical vapor deposition technique (MCVD) [2], [3]. Using Bi-doped silica-based fibers, laser operation in the spectral range of 1150 nm - 1225 nm was realised [4], [5]. Bi-ions incorporated into various glass hosts produce Bi-related active centers (BACs) with broadband emission and absorption spectra in the different regions of the near infrared (NIR) range [6].In contrast to rare-earth (RE) ions where optically active 4f electrons are shielded from the environment by the outermost 5s and 5p electrons, optical transitions in BACs are associated with the unshielded outer electron shell of a bismuth atom or ion. Therefore, the wavelengths of transitions and the structure of the energy levels depend considerably on the matrix of the host glass. The wavelength range of luminescence and optical amplification of Bi-doped glasses can be changed by varying the composition of the host glass. As a result, BDF lasers with different hosts work efficiently in various regions starting from 1150 nm -1550 nm [7]–[10]. Nowadays, BDFs have been successfully used for the creation of a number of devices including broadband optical amplifiers, continuous-wave (CW) and mode-locked fiber lasers, superluminescent sources, etc. in the spectral region of 1150 nm -1730nm [9,11,12].In particular, it was recently shown that the amplification process in the wavelength range 1600 nm -1800nm can be realized using Bi-doped high-$GeO_2$ silica-based fiber [11]. Light sources in 1600 nm -1800 nm wavelength range have a growing number of promising applications. The most important of these applications is medicine. For example, it has been reported that lasers at 1720 nm can selectively target lipid rich sebaceous glands in human skin tissue. The penetration depth in tissue can be easily adjusted by using the right wavelength, e.g., at 980 nm the tissue depth of damage is ~5nm compared to only ~3nm at 1700nm [13]. Unfortunately, the 1600 nm - 1800 nm wavelength band remains poorly covered by fiber lasers because of the absence of appropriate optical transitions of RE ions, sparking continued interest in the development of suitable sources. Lasers are typically designed to operate at a single wavelength. On the other hand, a dual-wavelength laser, which can produce two separate output wavelengths in a single structure, are useful for applications in sensing, imaging, and instrumentation [14]. Dual-wavelength and multi-wavelength lasers have been demonstrated by using different schemes, including cascaded fiber Bragg gratings (FBGs) [15], four-wave-mixing of dispersion shifted fiber [16], nonlinear polarization rotation [17], and various fiber-optic interferometers (FOIs) [18–26].

In this paper, we present a theoretical model of a dual-wavelength BDF laser based on cascaded cavities (Fig.1) and experimentally demonstrate its operation. The laser operates at two wavelengths, 1727 nm and 1729 nm, and operation can be switched from single wavelength to dual-wavelength by control of the pump conditions. Experimental results are in good agreement with simulations based on the theoretical model of a cascaded cavity laser

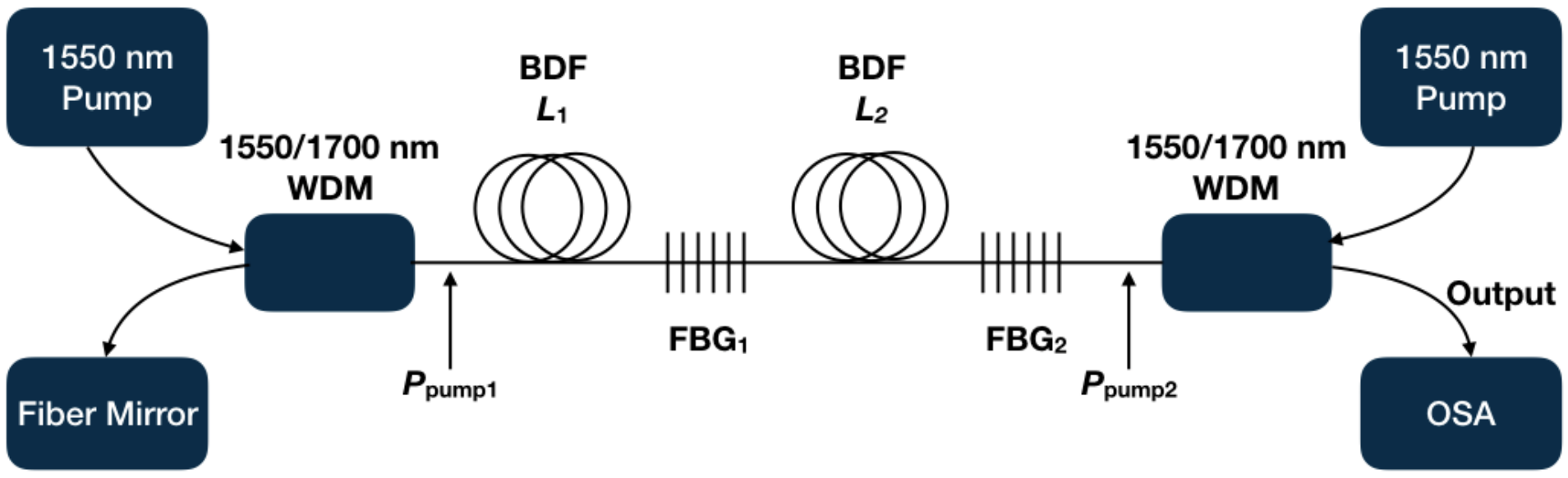

Fig. 1. The structure under investigation.BDF: bismuth-doped fiber; WDM: wavelength division multiplexer; OSA: optical spectrum analyzer.

## 2. Theoretical Analysis

The scheme of the dual-wavelength, cascaded cavity BDF laser is illustrated in Fig. 1. It comprises a broadband fiber mirror with reflectivity $R_0$ at the laser wavelengths which is placed on one end of BDF with alength$L_1$, and two fiber Bragg gratings ($FBG_1$ and $FBG_2$) separated with a second BDF with a length $L_2$ on the other end of the laser cavity. One of the gratings, $FBG_1$, has a peak reflectivity of $R_{L1}$ and is centered at $\lambda_1$; the second one, $FBG_2$, has a peak reflectivity $R_{L2}$ and is centered at $\lambda_2$. One of the cavities is defined by the fiber mirror, $FBG_1$, and the BDF with length $L_1$. The second laser cavity includes the same fiber mirror, $FBG_2$, and the cascade of BDFs with the lengths $L_1$ and$L_2$. We consider bidirectional pumping at a wavelength $\lambda_p$ = 1550 nm.

The two pumps are connected to laser cavities with wavelength division multiplexers (WDMs).The Bi-ions are considered as two-level active ions and can be described with well-known rate equations. The theoretical model is based on a set of rate equations and two radiation transport equations for the pump and signal intensities. Contrary to the boundary conditions used for traditional laser schemes, which consist of a single cavity with mirrors at both ends, the cascaded-cavity fiber laser requires separate boundary conditions at the fiber mirror (common reflector) and the two separate reflectors $FBG_1$ and $FBG_2$. It should be noted that the common cavity defined by BDF $L_1$ accounts for simultaneous propagation of both wavelengths $\lambda_1$ and $\lambda_2$. The rate equations are given as [27]

$$\frac{dN_2(z,t)}{dt} = -\left[W_e(z,t)+\frac{1}{\tau}\right]N_2(z,t)+\left[W_a(z,t)+W_p(z,t)\right]N_1(z,t), \tag{1}$$

$$N_T = N_1(z,t)+N_2(z,t), \tag{2}$$

where $N_T$ is the concentration the active particles in the host material, $N_2$ ($z$, $t$) and $N_1$ ($z$, $t$) are the population densities in the excited and the ground levels at the point $z$ along the length of the fiber at the moment $t$, $\tau$ is the lifetime of the excited level, and $W_{p,a,e}$($z$, $t$) are the light-induced pump ($p$), absorption ($a$), and emission ($e$) rates, respectively:

$$W_p = W_p^a P_p, \qquad W_e = W_p^e P_p + \sum_i W_{ASE}^{e(i)} P_{ASE}^{(i)}, \qquad W_a = \sum_i W_{ASE}^{a(i)} P_{ASE}^{(i)}, \tag{3}$$

$P_p = P_p^+ + P_p^-$ is the pump power and $P_{ASE}^{(i)} = P_{ASE}^{(i)+} + P_{ASE}^{(i)-}$ is the amplified spontaneous emission (ASE) power in the interval [$\lambda_i$-$\Delta\lambda$/2, $\lambda_i$+$\Delta\lambda$/2] , which propagate into the forward (+) and backward (-) directions. $\Delta\lambda$ is the step-size along the wavelength axis.

$$W_p^{a,e} = \frac{D\lambda_p}{hc\pi r_{co}^2}\sigma_{a,e}(\lambda_p), \qquad W_{ASE}^{a,e(i)} = \frac{\Gamma_i \lambda_i}{hc\pi r_{co}^2}\sigma_{a,e}(\lambda_i) \tag{4}$$

$$\text{where } D = \frac{\pi r_{co}^2}{A_p}, \qquad \Gamma_i = 1-\exp\left(-2\frac{r_{co}^2}{\omega_{0.i}^2}\right), \qquad \omega_{0,i} = r_{co}\left(0.65+1.619V_i^{-1.5}+2.876V_i^{-6}\right). \tag{5}$$

Here, $r_{co}$ is the fiber core radius, $V_i$ is the normalized frequency, $A_p$ is the entire pump cladding area, and $\sigma_{a,e}$ ($\lambda$) are the absorption ($a$) and emission ($e$) cross sections at wavelength $\lambda$. In steady-state, the time derivatives vanish ($d/dt$ = 0).The radiation transport equations for two level Bi-ions in solids, which describe the pump power and emission propagation along the length of the fiber, are similar to the radiation transport equations for a multilevel ions in solids such as, for example, $Er^{3+}$-doped solids [28]. For the two-level ion system, they have the following form:

$$\frac{dP_p(z)}{dz} = \{[-\sigma_a(\lambda_p)N_1(z) + \sigma_e(\lambda_p)N_2(z)]D - \alpha_p\}P_p(z), \tag{6}$$

$$\frac{dP_{ASE}^{(i)}(z)}{dz} = 2\frac{hc^2}{\lambda_i^3}\Delta\lambda\Gamma_i M_i \sigma_e(\lambda_i)N_2(z) + \{[\sigma_e(\lambda_i)N_2(z) - \sigma_a(\lambda_i)N_1(z)] - \alpha_i\}P_{ASE}^{(i)}(z). \tag{7}$$

where $M_i$ is the number of modes at the wavelength $\lambda_i$ and $\alpha_{p,i}$ are the loss parameters at the pump wavelength $\lambda_p$ ($p$) and the wavelengths $\lambda_i$($i$).

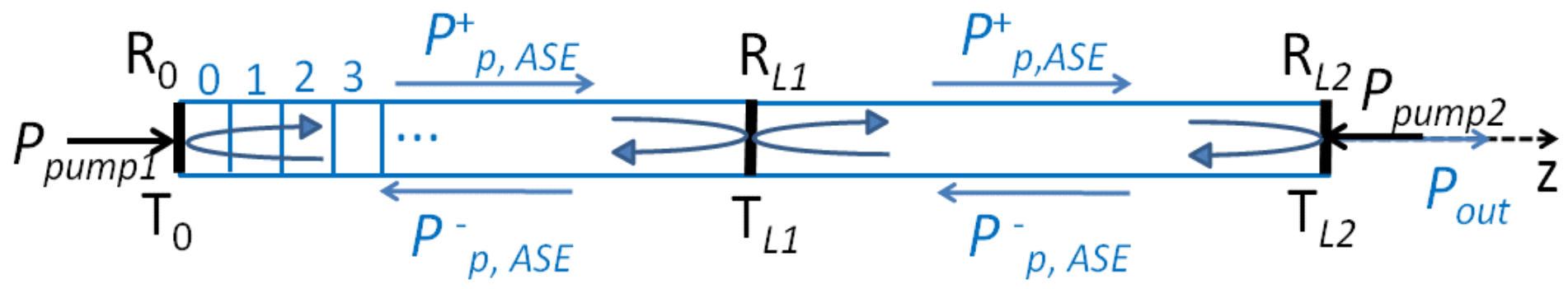

Fig. 2. Schematic of the numerical model.

Following to traditional approaches applied to fiber laser simulations, one can divide the BDFs with lengths $L_1$ and $L_2$ into $K_1$ and $K_2$ segments, respectively, providing as partial resolution $dz = L_{1,2}/K$. Each of these segments has its own population densities and optical powers. In our approach, they are equal to those at the left side of a segment when our iteration process moves from the left side to the right side of the scheme (Fig.2). They are equal to these values at the right side of the segment when our iteration process moves in opposite direction. A 2D grid over the length of the fiber and the wavelength range is used. The ASE power inside the laser cavity has to be reflected back to the cavity by the fiber mirror with reflection $R_0$ providing the following boundary condition $P_{ASE}^{+}(0) = R_0 P_{ASE}^{-}(0)$. The ASE power at the $FBG_2$ has to satisfy the following boundary condition $P_{ASE}^{-}(L_1 + L_2) = R_{L2} P_{ASE}^{+}(L_1 + L_2)$. At $FBG_1$ the ASE power signals traveling in the forward and backward directions will be partially reflected and transmitted. In our simulation scheme, all of the wavelengths reflected and transmitted at $FBG_1$ are taken into account in the iteration cycles. We assume that $R_0 + T_0 = 1$, $R_{L1} + T_{L1} = 1$, and $R_{L2} + T_{L2} = 1$.

## 3. Experimental Results

A single-mode BDF having a step-index refractive index profile was drawn from a preform fabricated by the modified chemical vapor deposition technique. The refractive index difference between the core and cladding of the preform was 0.06 (corresponding to ~50 mol.% of $GeO_2$). A detailed analysis of chemical glass composition was also performed by X-ray microanalysis. The cut off wavelength, core diameter and outer diameter of the fiber developed are 1.2 µm, 2 µm, and 125 µm, respectively. The total Bi concentration in the core of the fiber preform determined by means of electro-thermal atomization atomic absorption spectrometry (ETA-AAS) and inductively coupled plasma atomic emission spectroscopy (ICP-AES), was $\sim 10^{24}$ $m^{-3}$. It is well known that the concentration of the BACs is significantly lower than the total Bi content (e.g., [29] and references therein).We used an approach based on the measurement of large signal absorption saturation described in [30] to estimate the concentration of the BACs. It turns out that to be 10% with respect to the total concentration ($\sim 10^{23}$ $m^{-3}$); this is the value used in our simulations. The fiber core was made of Bi-doped silicate glass with high Ge content for achieving the bright luminescence in a spectral region around 1700nm. As shown in Fig. 3, it allows forming BACs with typical luminescence spectrum having a peak near 1700 nm and a width of ~200 nm. The absorption spectrum of the active fiber is also presented in Fig. 3. It consists of two broad distinctive bands centered at ~1400 nm and ~1650 nm. It should be noted that only the longer wavelength band can be used for pumping to obtain amplification between 1600 nm and 1800 nm [12, 31]. The peak absorption of both bands is relatively low due to small Bi concentration. The undesired concentration-assisted effects, particularly a growth of unsaturable absorption occurring for higher Bi concentrations, did not allow us to produce efficient Bi-doped fibers with high absorption level. Fig. 3 shows the spectral dependence of the unsaturable absorption in the fiber which was derived from the experimentally measured large signal absorption saturation curves at various wavelengths. Its value in the spectral range of interest is ~0.1 dB/m and weakly depends on wavelength. Despite the high concentration of $GeO_2$ in glass core the active fiber could successfully be spliced to standard SMF-28 fiber with a typical splicing loss of ~0.8 dB.

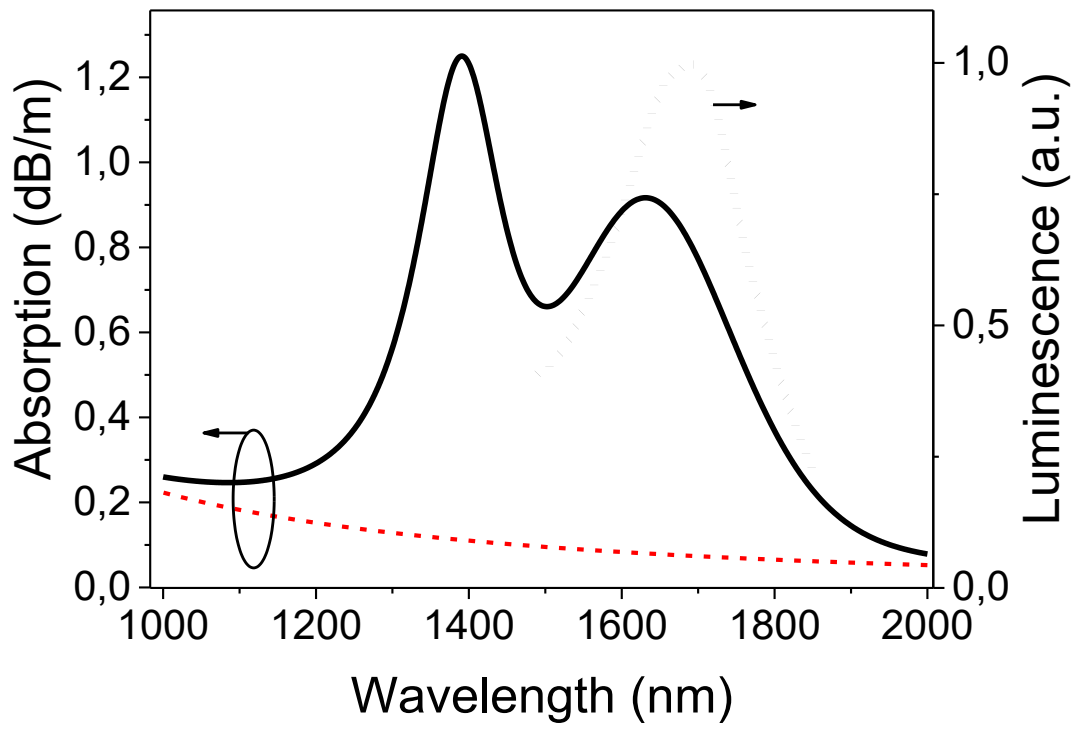

Fig. 3. Absorption (solid), unsaturable absorption (dashed) and luminescence (dotted) spectra of the used BDF.

Figure 1 shows the experimental setup of dual-wavelength cascaded cavity BDF laser. Two BDFs ($L_1 = L_2 = 20$ m) are spliced to SMF-28 fiber connectors and then used as the active gain medium. A bidirectional pumping scheme is employed along with the BDFs in a cascaded cavity configuration. The pumps come from a single external cavity laser at 1550 nm which is split by a coupler and the two separate signals are then amplified separately using two high power EDFAs. Such an arrangement allows to individually control the two pump powers to set the operating regime of the BDF laser. The 1550 nm pump and lasing signals are then coupled via 1550/1700 nm WDM couplers with insertion losses less than 1 dB. The 1550/1700 nm WDM on the left side of the experimental setup is capable of sustaining a maximum power $P_{pump1}$ of 1 W while the other WDM on the right side is capable of sustaining a maximum power $P_{pump2}$ of 300 mW. The cascaded laser cavity is defined by a fiber mirror on one end and FBGs on the other end. The broadband mirror is a fiber optic reflective-coated patch cable with a reflectivity of 90%. $FBG_1$ has a peak reflectivity of 95.2% at $\lambda_1 = 1727$ nm with a 3 dB bandwidth of 0.43 nm and $FBG_2$ has a peak reflectivity of 99.5% at $\lambda_2 = 1729$ nm with a 3dB bandwidth of 0.22 nm. The insertion loss of both FBGs is measured to be around 1 dB. The laser output is extracted via the 1550/1700 nm WDM coupler on the right side and measured using an OSA (AQ6317, ANDO) with 50pm resolution. No polarization controllers are used in the laser, as they were found to have negligible impact on laser performance.

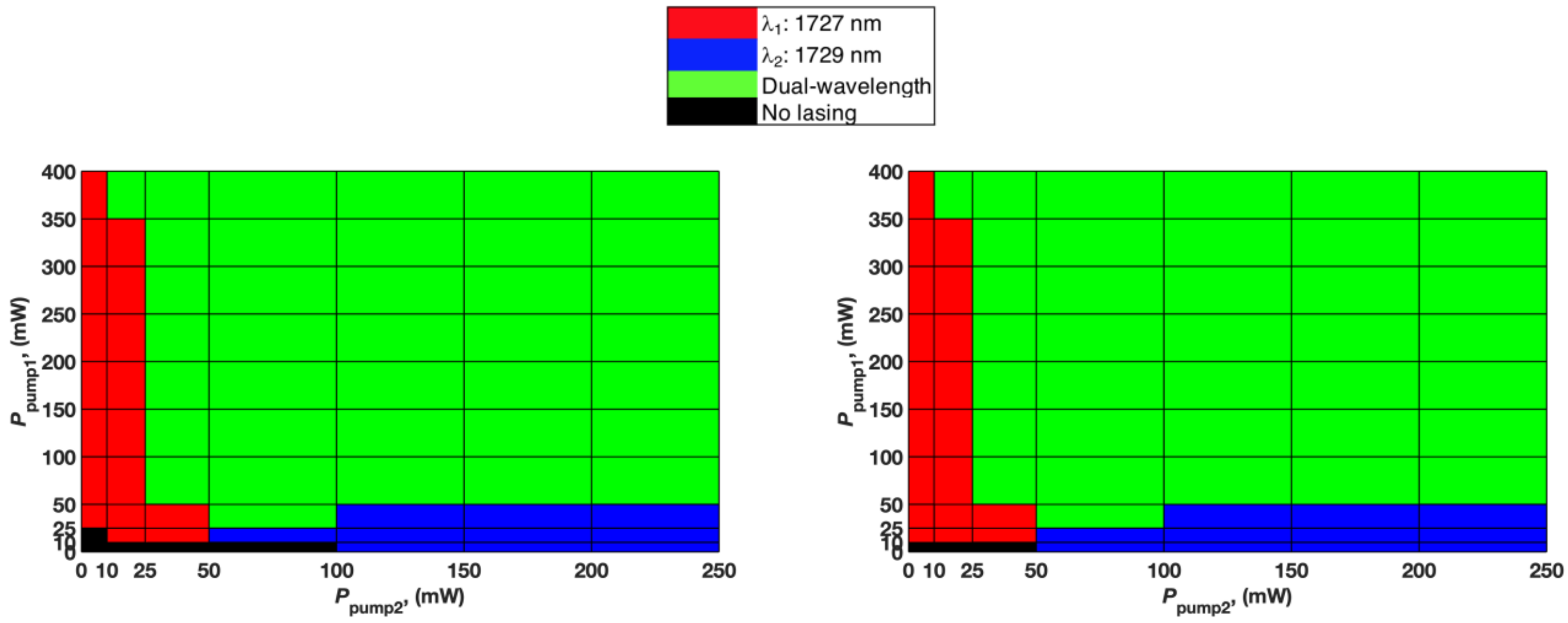

(a) (b)

Fig. 4. The contour maps based on (a) experiment and (b) theoretical model simulation of BDF laser oscillating regime against pump powers.

The operating BDF laser oscillation regimes in terms of output wavelengths vs. pump power are summarized and shown as contour maps in Fig. 4. The red, blue, green, and black regions in the contour maps represent the regimes where the BDF laser operates only at $\lambda_1 = 1727$ nm, $\lambda_2 = 1729$ nm, at both wavelengths, and below threshold, respectively. The measured contour map is shown in Fig. 4(a). It is obvious that alternate single wavelength operation as well as simultaneous dual-wavelength operation can be achieved by selecting different combinations of the bidirectional pump powers. For example, if $P_{pump1}$ is set to be off or at relatively low values (< 50 mW), only lasing at $\lambda_2 = 1729$ nm occurs when $P_{pump2}$ is properly arranged to exceed its threshold (largely > 50 mW). This is due to the fact that the cavity loss at $\lambda_1$ is too large to be overcome. Likewise, only lasing at $\lambda_1 = 1727$ nm occurs when the cavity loss at $\lambda_1$ is overcome by increasing $P_{pump1}$ and $P_{pump2}$ is set at the relatively low values (< 25 mW). Eventually, as both $P_{pump1}$ and $P_{pump2}$ get further increased, both $\lambda_1$ and $\lambda_2$ lasing conditions will be satisfied and dual-wavelength lasing is obtained. The simulated contour map is shown in Fig. 4(b); all simulation parameters are based on the experimental data. It is clear that simulated contour map shows good agreement with measurements. The only two mismatching lasing condition is when $P_{pump1}$ = 10 mW and $P_{pump2}$ is off and $P_{pump1}$ is off and $P_{pump2}$ = 50 mW. In both cases, no lasing occurs in experiment whereas simulations predict single wavelength operation. This discrepancy may be attributed to the differences in the extracted values of the BDF laser parameters used for simulation compared to the actual experiments.

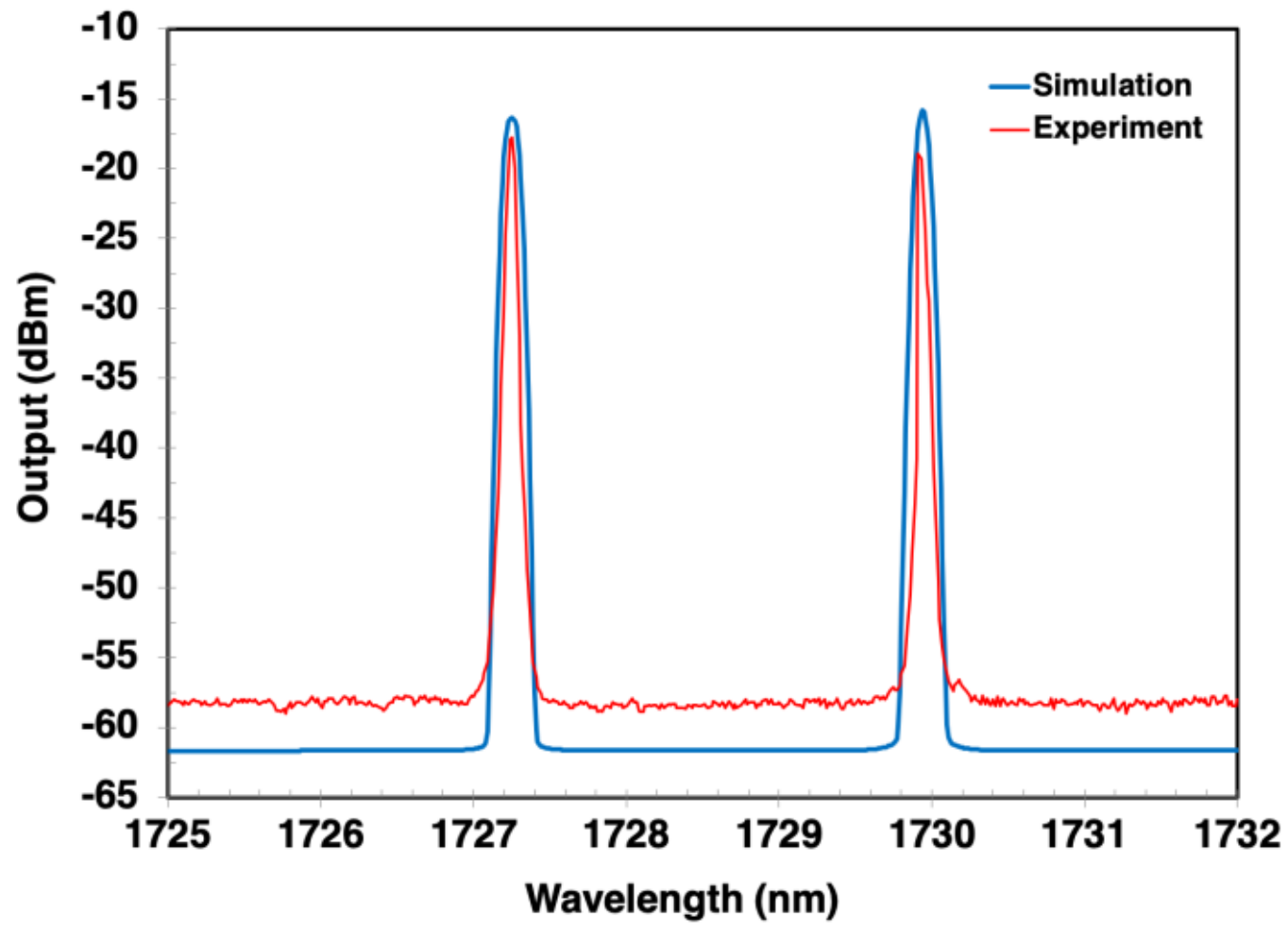

Fig. 5. Output spectrum of BDF laser during dual-wavelength operation when $P_{pump1}$= 100 mW and $P_{pump2}$ = 250 mW based on experiment (red) and theoretical model simulation (blue).

Representative BDF laser output spectrum for dual-wavelength operation with $P_{pump1}$ = 100 mW and $P_{pump2}$ = 250 mW is shown in Fig.5. The red trace shows the experimental BDF laser output spectrum. The output laser power is measured to be -17.8 dBm and -18.8 dBm at $\lambda_1$ and $\lambda_2$, respectively. It is seen that roughly equal laser powers at $\lambda_1$ and $\lambda_2$ (1 dB difference) can be achieved by appropriately setting $P_{pump1}$ and $P_{pump2}$, and the signal-to-noise ratio (SNR) is approximately 40 dB for both $\lambda_1$ and $\lambda_2$.The 3dB linewidths are both < 0.05nm (i.e., within the OSA resolution bandwidth). The blue trace shows the simulated BDF laser output spectrum and it shows good agreement with experiment. The output laser power is measured to be -16.3 dBm and -15.8 dBm at $\lambda_1$ and $\lambda_2$, respectively. Clearly, almost equal laser powers at $\lambda_1$ and $\lambda_2$ (< 0.5 dB difference) is observed with SNRs of 45 dB and 3 dB linewidths of < 0.1 nm (we use a wavelength resolution of 0.01 nm in the simulation).

## 4. Discussion

Based on the good agreement between our theoretical model and measurement results, we use numerical simulations to examine further the impact of varying the various parameters on laser performance. While simulations are conducted for the dual-wavelength, cascaded laser configuration, we show results for the output at 1727 nm (similar results are obtained at 1729 nm).

First, we consider the impact of varying the pump power, e.g., $P_{pump1}$, assuming $P_{pump2}$ is fixed at 250 mW, the lengths of the fiber are $L_1 = L_2$ = 20 m and the FBG reflectivities are 95.2% at $\lambda_1$=1727 nm ($FBG_1$), and 99.5% at $\lambda_2$ = 1729 nm ($FBG_2$). As one can see in Fig. 6(a), the output peak power linearly increases with the increase in one of the pump powers and reaches ~0.14 mW (-8.5 dBm) for $P_{pump1}$ = 1W. For $P_{pump1}$ = 100 mW as one can see from our simulations the output peak power at 1727 nm can reach ~0.025 mW (-16dBm). Experimentally the output peak power at 1727 nm was -17.8 dBm (Fig.5).

FBG reflectivities at 1727 nm and 1729 nm can be used for structure optimization. Let us consider a structure with the pump powers are $P_{pump1} = P_{pump2.}$ = 500 mW, and the BDF lengths are $L_1 = L_2$ = 20 m. It is important to emphasize that if we want to reach the dual-wavelength laser operation the reflectivities of FBGs have to be almost equal to each other. Let us consider the case when the reflectivities of the $FBG_1$ and $FBG_2$ are equal to each other. Fig. 6 (b) presents the dependence between the reflectivities of FBGs and the output peak power at the wavelength $\lambda_1$ = 1727 nm.As one can see in Fig. 6 (b), the reflectivities of $FBG_1$ and $FBG_2$ equal to ~25% are the best choice to reach the highest output peak power with this structure.

The lengths of BDFs ($L_1 = L_2$) can be considered as two parameters, which can be used for the laser structure optimization. The length of BDFs can be different or equal to each other. It is important to emphasize that all parameters of the laser structure such as BDF length, FBG reflectivity, the applied pump powers ($P_{pump1}$ and $P_{pump2}$) are tightly connected with each other on the way of structure optimization. Let us restrict ourselves by the structures with the equal lengths of BDFs ($L_1 = L_2$), since our experiments and our optimization of FBG reflectivities has been done for BDFs with the equal lengths. Fig. 6 (c) illustrates the dependence between the length of BDFs ($L_1 = L_2$) and the output peak power, $P_{out}$, at the wavelength $\lambda_1 = 1727$ nm when $P_{pump1} = P_{pump2.} = 500$ mW and the reflectivity of FBG1 and FBG2 is 25% and 29%, respectively. In this simulation, the BDF lengths equal to 20 m is seen to be the best choice. Indeed, in longer than ~20m fibers the laser signal can be reabsorbed. Shorter than ~20 m fibers cannot completely absorb the pump power. As a result they have the relatively low output power. The future optimization of the scheme is possible.

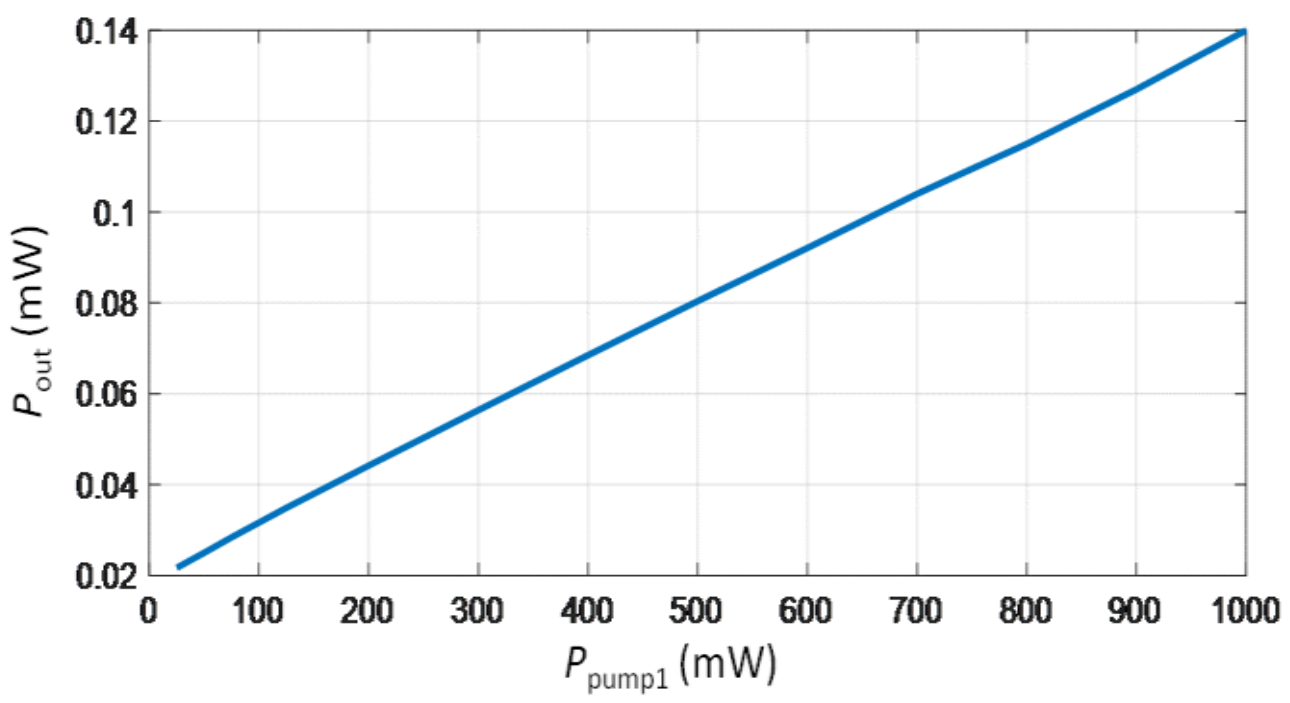

(a)

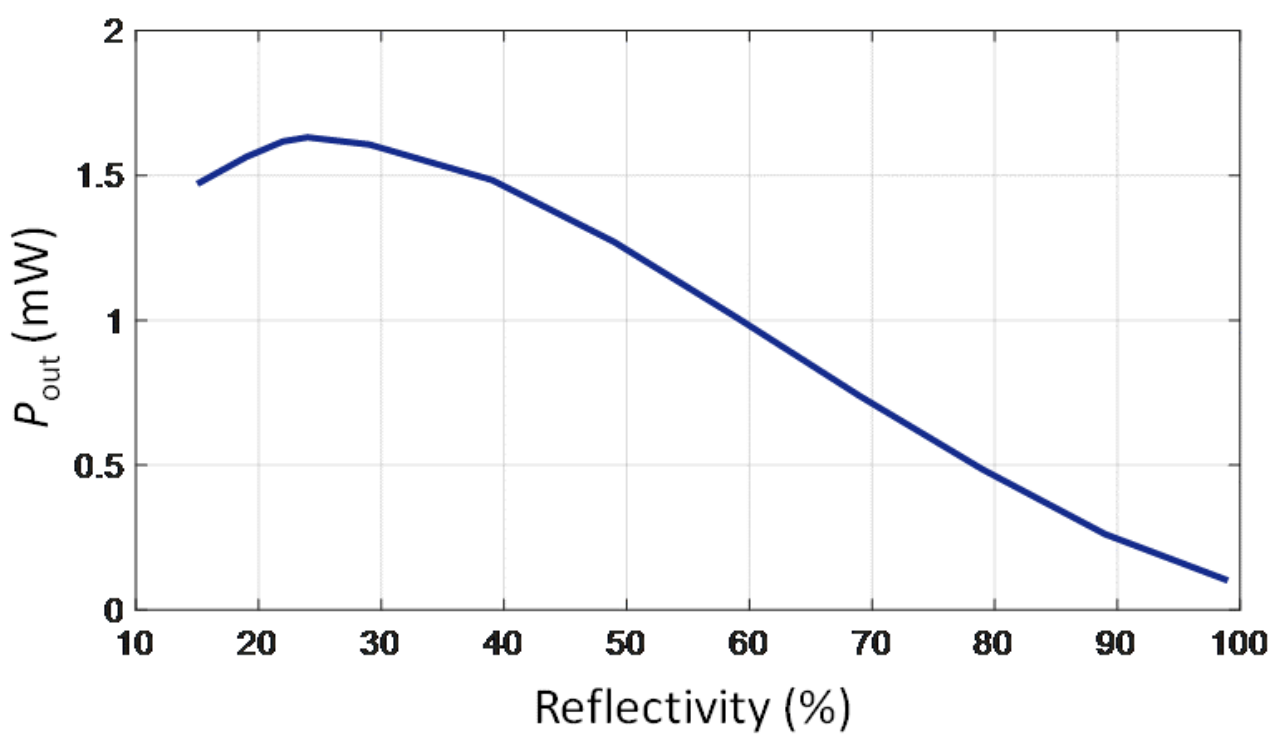

(b)

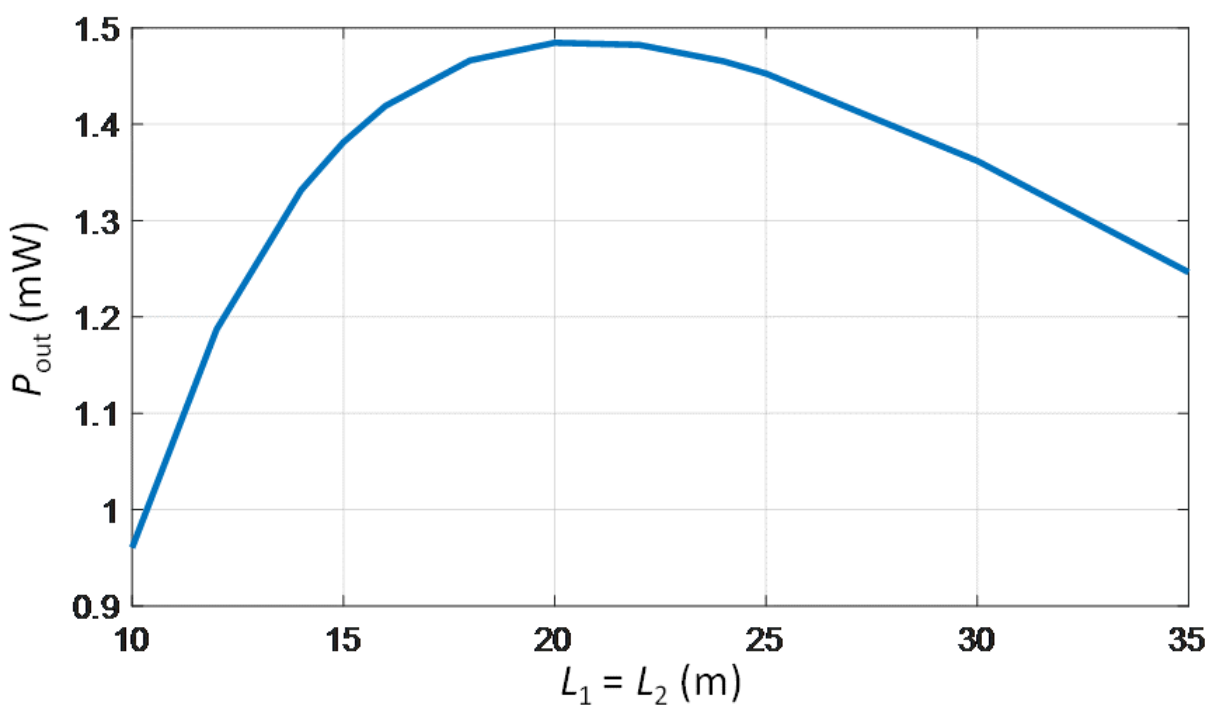

(c)

Fig. 6. (a)The output peak power at the wavelength 1727 nm as a function of the pump power. $P_{pump2} = 250$ mW. (b)The output peak power at the wavelength 1727 nm as a function of the FBG reflectivity. The reflectivity of FBG1 and FBG2 are equal each other. (c) The output peak power at the wavelength 1727 nm as a function of the BDF lengths.

The low efficiency of the laser is due to several reasons. First, the BDF has a very small core diameter which results in higher splicing losses to SMF. As a result, the total insertion losses of signal as well as pump in the realized scheme may to be 3-4 dB for one pass. One of the possible ways for optimization is to fabricate a fiber having refractive index profile with pedestal geometry. In addition, the BDF is also characterized by a high level of the unsaturable absorption. It also influences the output characteristics of the dual-wavelength laser. It should be noted that a similar problem was encountered at the initial stages when developing fibers doped with rare-earth ions. This was successfully solved by improving the manufacturing technology and choosing the appropriate chemical composition of the glass matrix [32]. We believe that the similar approaches could be worked in case of Bi-doped fibers. However, it is beyond the scope of our paper.

## 5. Summary

BDFs have unique optical properties, which allow them to be one of most promising gain media in a spectral region of 1600 nm – 1800 nm. The interest to BDFs has been generated by the development of the first CW and mode-locked lasers, amplifiers, broadband fiber sources operating in this wavelength range. The development of multi-wavelength lasers in the 1700 nm wavelength range is a new step towards the application of this new active medium. Our theoretically analyzed and experimentally demonstrated dual-wavelength cascaded cavity BDF laser pumped at 100 mW and 250 mW has reached ~ –17 dBm output peak power at 1727 nm and 1729 nm. As we see from our simulations the scheme can be optimized. The higher output peak powers at 1727 nm and 1729 nm can be reached by optimization of the lengths of BDFs, as well as the FBG reflectivity and the pump powers. The properly optimized scheme and properly arranged pump powers permit one to reach the output peaks with the equal powers. The dual-wavelength operation can by switched to the single wavelength operation by the proper choice of the pump powers. The future development of BDFs can open new possibility in this useful wavelength range.

## Acknowledgements

The authors wish to thank the anonymous reviewers for their valuable suggestions. The research was supported in part by the Russian Foundation for Basic Research (grant number 18-32-20003)